\documentclass{webofc}
\usepackage[varg]{txfonts}
\usepackage{slashed}
\usepackage{hyperref}

\newcommand \ml {m_l}
\newcommand \ms {m_s}
\newcommand \U {\mathcal{U}}
\newcommand \av[2] {\left\langle{#1}\right\rangle_{#2}}
\newcommand \pbp {\bar{\psi}\psi}
\newcommand \cum[1] {\mathbb{K}_{#1}}
\newcommand \lda {\lambda}
\newcommand \ru[1] {\rho_U(\lda_{#1})}
\newcommand \pu[2] {P_U(\lda_{#1};{#2})}
\newcommand \pn {P_n(\lda)}
\newcommand \fn {f_n(z)}
\newcommand \gn {g_n(\lda)}

\newcommand \tc {T_c}
\newcommand \hlda {\hat{\lda}}
\newcommand \hp[1] {\hat{P}_{#1}(\hlda)}
\newcommand \hml {\hat{m}_l}
\newcommand \hgn {\hat{g}_n(\hlda)}
\newcommand \D {\slashed{D}}
\begin{document}
\title{Microscopic Origin of Criticality at Macroscale in QCD Chiral Phase Transition}
\author{
\firstname{Heng-Tong} \lastname{Ding}\inst{1}\fnsep\thanks{\email{hengtong.ding@ccnu.edu.cn}} \and
\firstname{Wei-Ping}  \lastname{Huang}
    \inst{1}\fnsep\thanks{\email{huangweiping@mails.ccnu.edu.cn}} \and
\firstname{Swagato}   \lastname{Mukherjee}
	\inst{2}\fnsep\thanks{\email{swagato@bnl.gov}} \and
\firstname{Peter}     \lastname{Petreczky}
	\inst{2}\fnsep\thanks{\email{petreczk@bnl.gov}}
}
\institute{
Key Laboratory of Quark \& Lepton Physics (MOE) and Institute of Particle Physics, \\ Central China Normal University, Wuhan 430079, China 
\and
Physics Department, Brookhaven National Laboratory, Upton, NY 11973, USA
}
\abstract{%
We reveal that the criticality of the chiral phase transition in QCD at the macroscale arises from the microscopic energy levels of its fundamental constituents, the quarks. We establish a novel relation between cumulants of the chiral order parameter (i.e., chiral condensate) and correlations among the energy levels of quarks (i.e., eigenspectra of the massless Dirac operator), which naturally leads to a generalization of the Banks-Casher relation. Based on this novel relation and through (2+1)-flavor lattice QCD calculations using the HISQ action with varying light quark masses in the vicinity of the chiral phase transition, we demonstrate that the correlations among the infrared part of the Dirac eigenspectra exhibit same universal scaling behaviors as expected of the cumulants of the chiral condensate. We find that these universal scaling behaviors extend up to the physical values of the up and down quark masses.
}
\maketitle
\vspace{-3ex}
\section{Introduction}
\label{sec:intro}
One of the central goals in heavy-ion collision experiments is to search for universal signatures of criticality in the phase diagram of strong-interaction matter governed by QCD. Various macroscopic observables for this purpose have been measured~\cite{Luo:2017faz}. Even though these macroscopic critical behaviors are finally established, how such criticality at the macroscale arises from the fundamental constituents and interactions of QCD remains unanswered. 

Moreover, theoretical pursuit of the microscopic origin of the criticality in QCD chiral phase transition is difficult to proceed. The universal feature of the second order chiral phase transition in QCD with massless up and down quarks and a strange quark with physical mass~\cite{Pisarski:1983ms,HotQCD:2019xnw,Ding:2020xlj} is that macroscopic quantities related to the order parameter follow certain power law scaling behaviors that are uniquely characterized by the dimensionality and global symmetries of the system, irrespective of the details of its microscopic degrees of freedom and interactions. Consequently, although a simplified effective theory possessing the same dimensionality and global symmetries of QCD can be used to understand macroscopic properties of a system near the transition region, it  cannot unveil properties at the microscale due to ignorance of microscopic complexities of QCD in such effective theories. In this work, we will present a first lattice QCD based understanding of connections between the universal features at macroscopic and microscopic scales of QCD. Detailed information about this work can be found in Ref.~\cite{Ding:2023oxy}.
\section{Theoretical developments}
\label{sec:theor-develop}
For (2+1)-flavor QCD with degenerate light up ($u$) and down ($d)$ quarks having masses $\ml=m_u=m_d$ and a heavier strange quark with physical mass $\ms$, for simplicity we develop the main theoretical idea by just considering QCD with degenerate light quarks in this section.

We start with a generating functional defined as
\vspace{-1ex}
\begin{align}
  \mathbb{G}(\ml;\epsilon) = \ln \av{ \exp \left\{ - \ml \pbp(\epsilon) \right\} } {0} \,,
\label{eq:gf1}
\end{align}%
where a probe operator $\pbp(\epsilon)\equiv 2\text{Tr}(\D[\U] + \epsilon)^{-1}$ is introduced to probe the system in the chiral limit with a valence quark mass $\epsilon$ used to facilitate the evaluation of it. Here $\D[\U]$ is the massless Dirac operator for a given background SU(3) gauge field $\U$; trace is applied over the color, spin and space-time indices; $\av{\cdot}{0}$ denotes a expectation value with respect to the QCD partition function in the chiral limit $Z(0)=\int \exp\{-S[\U,0]\} \mathcal{D}[\U]$, where $S[\U,\ml] = S_g[\U] + \bar{\psi}\D[\U]{\psi} + \ml\pbp$ is the Euclidean-time QCD action, $S_g[\U]$ is the pure gauge action and $\pbp = \bar\psi_u\psi_u + \bar\psi_d\psi_d$.
Then the $n$th order cumulants, $\cum{n}$, of the chiral order parameter $\pbp(\ml)$ can be obtained from derivatives of above generating functional
\vspace{-1ex}
\begin{align}
  \cum{n} \left[ \pbp \right]= \frac{T}{V} \, (-1)^n \left. \frac{\partial^n \mathbb{G}(\ml;\epsilon)}{\partial \ml^n} \right\vert_{\epsilon=\ml} \,,
\label{eq:kn1}
\end{align}%
where $T$ is the temperature and $V$ is the spatial volume of the system. Recognizing $\av{\mathcal{O}}{} = \av{ \mathcal{O} \exp\{ - \ml \pbp(\ml) \} } {0}/\av{\exp\{ - \ml \pbp(\ml) \} } {0}$ and $Z(\ml)/Z(0)= \av{ \exp\{ - \ml \pbp(\ml) \} } {0}$, it is easy to see that $\cum{n}$ are the standard cumulants of $\pbp(\ml)$; e.g., $\cum{1}\left[ \pbp \right]=T\langle\pbp(\ml)\rangle/V$, $\cum{2}\left[ \pbp \right]=T\langle[\pbp(\ml)-\langle\pbp(\ml)\rangle]^2\rangle/V$, $\cum{3}\left[ \pbp \right]=T\langle[\pbp(\ml)-\langle\pbp(\ml)\rangle]^3\rangle/V$, etc.

The eigenvalues $\lda_j[\U]$ of $\D[\U]$ give the energy levels of a massless quark in given background $\U$, and the probe operator is further related to eigenspectra $\ru{} = \sum_j \delta(\lda-\lda_j)$ as $\pbp(\epsilon)\equiv 2\text{Tr}(\D[\U] + \epsilon)^{-1} = 2 \sum_j (i\lda_j+\epsilon)^{-1}=\int_0^{\infty}d\lda~4\epsilon\ru{}/(\lda^2 + \epsilon^2)$. Thus, Eq.~\ref{eq:gf1} becomes
\vspace{-2ex}
\begin{align}
  \mathbb{G}(\ml;\epsilon) = \ln \av{ \exp \left\{ - \ml \int_0^\infty \negthickspace \negthickspace \pu{}{\epsilon} d\lda  \right\} } {0} \,,
  \quad\text{where}\quad
  \pu{}{\epsilon} \equiv \frac{4\epsilon\ru{}}{\lda^2 + \epsilon^2} \,.
\label{eq:gf2}
\end{align}
Hence through derivatives shown as Eq.~\ref{eq:kn1} it is straightforward to obtain our main theoretical result connecting the cumulants of the order parameter to the $n$-point correlations of the quark energy levels $\ru{}$:
\vspace{-2ex}
\begin{align}
  \cum{n} [\pbp] = \int_0^\infty \negthickspace \negthickspace \pn d\lda  \,,
\label{eq:kn2}
\end{align}%
where $P_1(\lda)=K_1[\pu{}{\ml}]$ for $n=1$, and for $n\geq2$
\vspace{-1ex}
\begin{align}
  P_n(\lda)=\int_0^\infty  K_1\big[\pu{}{\ml}, \pu{2}{\ml}, \dotsc, \pu{n}{\ml} \big] ~\prod_{i=2}^{n}d\lda_i  \,.
\label{eq:pn}
\end{align}%
Here $K_1$ is the first order joint cumulant of $n$ variables.

In the vicinity of the chiral transition of the staggered lattice QCD at nonvanishing lattice spacings~\cite{Kilcup:1986dg,Ejiri:2009ac,Clarke:2020htu} the following relation is expected~\cite{Ding:2023oxy} 
\vspace{-2ex}
\begin{align}
  \cum{n}[\pbp] = \int_0^\infty \negthickspace \negthickspace \pn d\lda \sim \ml^{1/\delta-n+1} f_n(z) \,.
\label{eq:scaling}
\end{align}
Here the scaling variable $z \propto z_0 \ml^{-1/\beta\delta}(T-\tc)/\tc$, where $\tc$ is the chiral phase transition temperature and $z_0$ is a scale parameter; both are system specific. $\beta$ and $\delta$ are the universal critical exponents, and $f_{n+1}(z) = (1/\delta-n+1) f_{n}(z) - z f_{n}^{\prime}(z) / {\beta\delta}$ are  universal scaling functions belonging to a 3-d O(2) universality class. Here, $n\ge1$ and the superscript prime denotes derivative with respect to $z$. In our work we adopted $\beta=0.349$, $\delta=4.78$ and for consistency the scaling functions $f_n(z)$ of the O(2) universality class determined in Refs.~\cite{Engels:2001bq,Ejiri:2009ac}. 

Eq.~\ref{eq:scaling} indicates that universal scaling properties of the macroscopic observables $\cum{n}[\pbp]$ arise from the correlations among the microscopic energy levels $\pn$. To elucidate this point, consider $\ml\to0$. Recognizing $\pu{}{\epsilon\to0} = 2\pi \ru{} \delta(\lda)$ from Eq.~\ref{eq:gf2}, one finds $\lim_{\ml\to0}P_1(\lambda)=2\pi K_1[\ru{}]\delta(\lda)$ and $\lim_{\ml\to0} \pn = (2\pi)^n K_1[\ru{}, [\rho_U(0)]^{n-1}] \delta(\lda)$ for $n\geq2$ (from Eq.~\ref{eq:pn}). Noting that $K_1$ of $n$ identical variables is equivalent to $\cum{n}$, Eq.~\ref{eq:kn2} in the chiral limit thus becomes a novel generalization of Banks-Casher relation~\cite{Banks:1979yr} expressed as follows:
\vspace{-2ex}
\begin{align}
  \lim_{\ml\to0} \cum{n} [\pbp] = (2\pi)^n \cum{n} [\rho_U (0)] \,.
\label{eq:genBC}
\end{align}
\vspace{-3ex}

In the chiral limit and around $\tc$, according to Eq.~\ref{eq:genBC} the universal scaling behavior manifested in $\cum{n}[\pbp]$ must arise from the universal behaviors of the $\lambda$-independent $\cum{n}[\rho_U(0)]$. Thus, it is natural to expect for small $\ml$ within the scaling window the critical scaling of $\cum{n}[\pbp]$ in Eq.~\ref{eq:scaling} arises from the universal behaviors of the amplitudes of $\pn$ at the infrared but not from its system-specific $\lambda$ dependence; i.e., $\pn = \ml^{1/\delta-n+1} \fn \gn$, where $\gn$ are non-universal functions encoding the properties of specific system under consideration.
\section{Results}
\label{sec:results}
To check our conjecture for $\pn$ numerically, (2+1)-flavor lattice QCD calculations with the HISQ action were performed for $T\in [135, 176]$ MeV, where $\ms$ was fixed to its physical value with a varying $\ml = \ms/27, \ms/40, \ms/80, \ms/160$ that correspond to Goldstone pion masses $m_{\pi}\approx$ 140, 110, 80, 55 MeV. Details about the lattice setup can be found in Ref.~\cite{Ding:2023oxy}.

Owing to Eq.~\ref{eq:genBC} we expect the relevant infrared energy scale is $\lda\sim\ml$ for small values of $\ml$. It is natural to express all quantities as functions of the dimensionless and renormalization group invariant $\lda/\ml$ and following notations are used hereafter: 
\vspace{-1ex}
\begin{align}
\begin{split}
  & \hlda = \lda / \ml \,, \quad 
  \hml = \ml / \ms \,, z=z_0\hml^{-1/\beta\delta}(T-\tc)/\tc\,,\\
  & \hp{n} = \ms^{n+1} \hml P_n(\lda) / \tc^4 \,, 
  \quad \text{and} \, \quad 
  \mathbb{\hat K}_{n} [\pbp] = \ms^n \cum{n} [\pbp] / \tc^4 = \int_0^\infty \negthickspace \negthickspace \hp{n} d\hlda \,.
\end{split}
\label{eq:hatdef}
\end{align}
The system-specific parameters $\tc=144.2(6)$ MeV and $z_0=1.83(9)$ needed below to obtain $f_n(z)$ were taken from Ref.~\cite{Clarke:2020htu}, where 3-dimensional O(2) scaling fits were carried out for the same lattice ensembles but using an entirely different macroscopic observable, namely the $\ml$ dependence of the static quark free energy.

\begin{figure*}[!htp]
  \centering
  \setlength{\abovecaptionskip}{0ex}
    \includegraphics[width=0.3\textwidth,height=0.15\textheight]{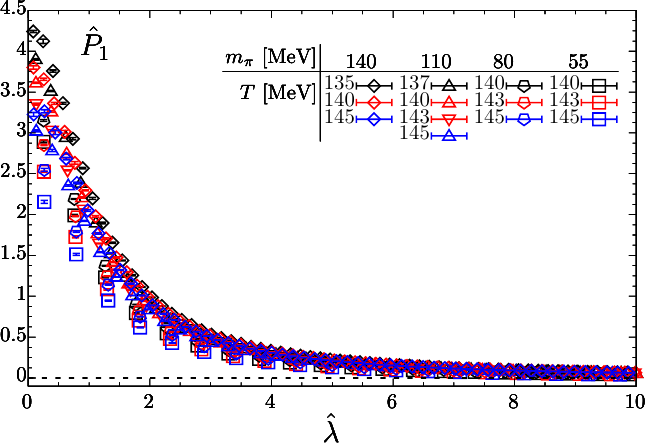}
    \includegraphics[width=0.3\textwidth,height=0.15\textheight]{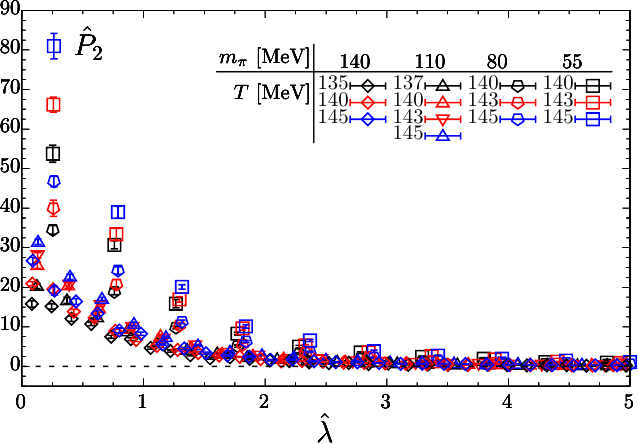}
    \includegraphics[width=0.3\textwidth,height=0.15\textheight]{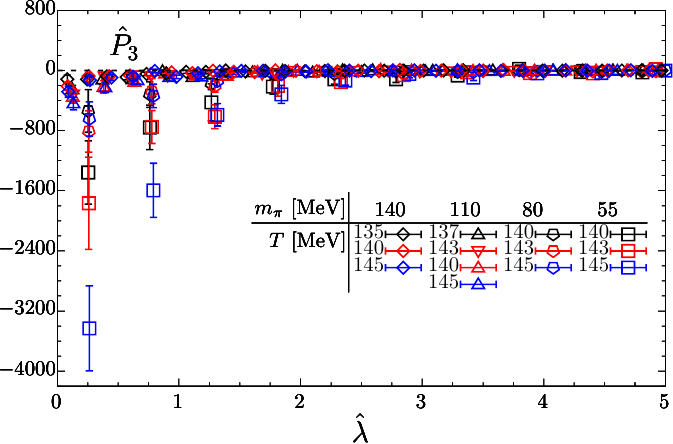}
  	\caption{$\hp{1}$ (left), $\hp{2}$ (middle) and $\hp{3}$ (right) for 135~MeV~$\le T \le$~145~MeV and 55~MeV~$\le m_\pi \le$~ 140~MeV.}
	\label{fig:P123}
\end{figure*}

Fig.~\ref{fig:P123} shows $\hp{n}$ for $n\leq 3$ as a function of $\hlda$ in the proximity of $\tc$. $\hp{n}$ rapidly vanishes for $\hlda\gtrsim1$, and the regions where $\hp{n}\ne0$ get smaller with increasing $n$. This reinforces that the relevant infrared energy scale turns out to be $\hlda\sim1$. In this infrared region $\hp{n}$ at a fixed $T$ shows clear dependences on $\ml$, which becomes stronger for increasing $n$. The form of $\ml$ dependence of $\hp{n}$ also changes with varying $T$. Expectedly, our results become increasingly noisy with increasing $n$ and decreasing $\ml$. With our present statistics we cannot access correlation functions with $n>3$, particularly for smaller $\ml$.

\begin{figure*}[!htp]
  \centering
  \setlength{\abovecaptionskip}{0ex}
    \includegraphics[width=0.3\textwidth,height=0.15\textheight]{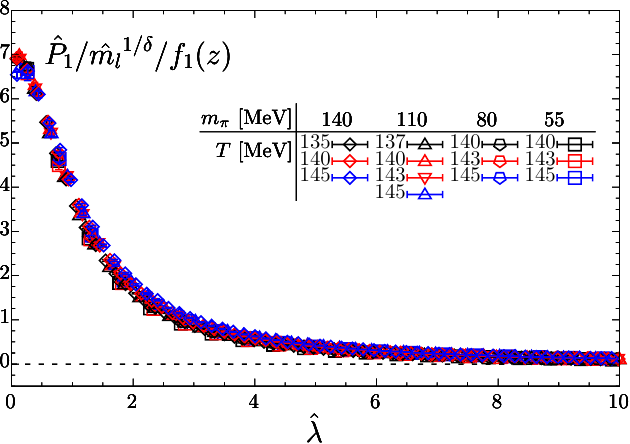}
    \includegraphics[width=0.3\textwidth,height=0.15\textheight]{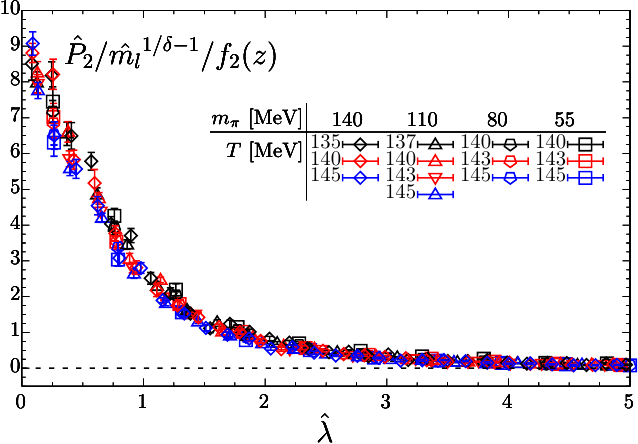}
    \includegraphics[width=0.3\textwidth,height=0.15\textheight]{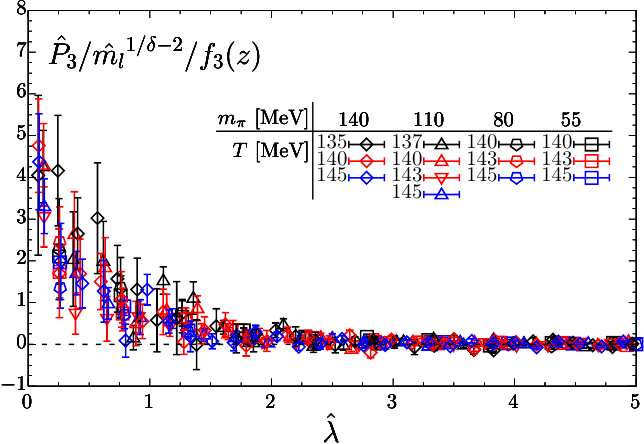}
  	\caption{$\hp{n}$ in Fig.~\ref{fig:P123} rescaled by $\hml^{1/\delta+1-n}f_n(z)$ for $n=1$ (left), $n=2$ (middle) and $n=3$ (right).}
 \label{fig:RescaledP123}
\end{figure*}

The $\ml$ and $T$ dependence of $\hp{n}$ shown in Fig.~\ref{fig:P123} can be understood in terms of the 3-d O(2) scaling properties. Once the $\hp{n}$ are rescaled with respective $\hml^{1/\delta+1-n}f_n(z)$ the data in Fig.~\ref{fig:P123} magically collapse onto each other (see Fig.~\ref{fig:RescaledP123}). Thus, our expectations for $\pn$ are clearly borne out in Fig.~\ref{fig:RescaledP123}, namely $\hp{n} = \hml^{1/\delta-n+1} \fn \hgn$.
Here $\hgn$ characterize the system specific of the $n$th order energy-level correlations. To satisfy our generalized Banks-Casher relations of Eq.~\ref{eq:genBC} the $\hgn$ must also satisfy $\lim_{V\to\infty} \lim_{a\to0} \lim_{\ml\to0} \hgn \to \delta(\hlda)$, such that $\cum{n}[\pbp]$ has the correct scaling behavior in $(T-T_c)/T_c$. It is noteworthy that the physical QCD with $m_\pi\approx$~140~MeV also shows same scaling for $T\in [135,145]$ MeV. Outside this temperature window we do not observe the scaling.
\vspace{-1ex}
\vspace{-3ex}
%
%

% \section*{Acknowledgement}
% \label{sec:ackonw}
%
\noindent \textbf{Acknowledgement.}
We thank Yu Zhang for early involvement, Sheng-Tai Li for technical support, Jacobus Verbaarschot and members of HotQCD collaboration for discussions. This work was supported by the National Key Research and Development Program of China under Contract No. 2022YFA1604900; the NSFC under grant No.~12293064, No.~12293060 and No.~12325508; the U.S. Department of Energy, Office of Science, Office of Nuclear Physics through Contract No.~DE-SC0012704 and within the framework of SciDAC award Fundamental Nuclear Physics at the Exascale and Beyond. Simulations were carried out on Nuclear Science Computing Center at CCNU, Wuhan supercomputing center and facilities of the USQCD Collaboration funded by the Office of Science of the U.S. Department of Energy.
\bibliography{ref.bib}

\begin{thebibliography}{10}

\bibitem{Luo:2017faz}
X.~Luo, N.~Xu, Nucl. Sci. Tech. \textbf{28}, 112 (2017), \texttt{1701.02105}

\bibitem{Pisarski:1983ms}
R.D. Pisarski, F.~Wilczek, Phys. Rev. D \textbf{29}, 338 (1984)

\bibitem{HotQCD:2019xnw}
H.T. Ding et~al. (HotQCD), Phys. Rev. Lett. \textbf{123}, 062002 (2019), \texttt{1903.04801}

\bibitem{Ding:2020xlj}
H.T. Ding et~al., Phys. Rev. Lett. \textbf{126}, 082001 (2021), \texttt{2010.14836}

\bibitem{Ding:2023oxy}
H.T. Ding et~al., Phys. Rev. Lett. \textbf{131}, 161903 (2023), \texttt{2305.10916}

\bibitem{Kilcup:1986dg}
G.W. Kilcup, S.R. Sharpe, Nucl. Phys. B \textbf{283}, 493 (1987)

\bibitem{Ejiri:2009ac}
S.~Ejiri et~al., Phys. Rev. D \textbf{80}, 094505 (2009), \texttt{0909.5122}

\bibitem{Clarke:2020htu}
D.A. Clarke et~al., Phys. Rev. D \textbf{103}, L011501 (2021), \texttt{2008.11678}

\bibitem{Engels:2001bq}
J.~Engels et~al., Phys. Lett. B \textbf{514}, 299 (2001), \texttt{hep-lat/0105028}

\bibitem{Banks:1979yr}
T.~Banks, A.~Casher, Nucl. Phys. B \textbf{169}, 103 (1980)

\end{thebibliography}

\end{document}